# Spatial and temporal evolutions of blue-core helicon discharge driven by planar antenna with concentric rings

Chao Wang[1,2], Lei Chang[1*], Ling-Feng Lu[2], Shunjiro Shinohara[3], Zhi-De Zeng[2], Ilya Zadiriev[4], Elena Kralkina[4], Zhi Li[1,2], Shi-Jie Zhang[1], Zi-Chen Kan[1], Ye Tao[1], Ding-Zhou Li[1],

[1]State Key Laboratory of Power Transmission Equipment Technology, School of Electrical Engineering, Chongqing University, Chongqing, 400044, China

[2]Southwestern Institute of Physics, Chengdu, 610041, China

[3]Institute of Engineering, Tokyo University of Agriculture and Technology, Tokyo 184-8588, Japan

[4]Physical Electronics Department, Faculty of Physics, Lomonosov Moscow State University, GSP-1, Leninskie Gory, Moscow, 119991, Russian Federation

Corresponding Email: leichang@cqu.edu.cn

**Abstract**

The spatial and temporal evolutions of blue-core helicon discharge driven by a planar antenna with four concentric rings are explored on the Linear Experimental Advanced Device (LEAD). The discharge experiences distinct density jumps from E mode to H mode, W mode, and blue-core mode, when RF input power increases. This is similar to previous observations using other typical helicon antennas; however, this special antenna could drive modes of even higher levels for which the blue-core plasma column is actually hollow in radius, i.e. peaking off-axis, which was not presented before. The column shows counterclockwise rotation for blue-core mode and clockwise rotation for non-blue-core mode. The reason could be attributed to the radial electric field differenceses for both modes which reverses the rotation direction via $\boldsymbol{E} \times \boldsymbol{B}$ drive. Moreover, the centrifugal instability of blue-core helicon plasma is computed using a two-fluid flowing plasma model. It shows that the instability is strong for small axial wave number but becomes weak for large axial wave number. Perturbed density peaks at radius of 0.045 m, while the equilibrium density gradient peaks at radius of 0.055 m. The coincidence of their radial locations suggests that it is a resistive drift mode driven by density gradient. The blue-core mode weakens once the magnetic field or flow rate exceeds the threshold value. Increasing power further leads to a smoother plasma density gradient. The electron temperature profiles decrease with increased power, and the radial gradient of the electron temperature inside the core is smaller as the magnetic field changes. To our best knowledge, it is the first detailed characterization of blue-core helicon plasma driven by planar antenna, especially in terms of azimuthal rotation and centrifugal instability.

## 1. Introduction

Since Boswell first described helicon plasma in 1970, it has been actively investigated around the world [1]. The helicon plasma source is a type of radio frequency (RF) plasma source that can deposit power into plasma through an antenna. The frequency of helicon plasma lies between ion cyclotron frequency and electron cyclotron frequency [2-4]. Helicon plasma sources are notable for high density (the density can reach $10^{20}\,\mathrm{m^{-3}}$), high ionization (near full ionization rate inside the core), and high power efficiency compared to inductively coupled plasma (ICP) and capacitively coupled plasma (CCP). Therefore, helicon plasma sources have received widespread world attention and have been extensively applied in various fields such as material processing [5], space propulsion [6], semiconductor manufacturing [7], and nuclear fusion [8, 9]. In addition, helicon plasma sources are used as a platform for the study of fundamental physics [10, 11]. Despite extensive research on helicon plasma over many years, the underlying physical processes responsible for the formation of blue-core plasma [12, 13], effective power deposition [14-17], discharge in multiple wave modes [18], and mode transition [19] remain incompletely comprehended.

The discharge in the conventional helicon source setup demonstrates a distinctive feature known as mode transition, which is marked by a step rise in electron density as the radio frequency power or magnetic field increases. This discharge mode encompasses capacitively coupled (E) mode at first, and then shifts into inductively coupled (H) mode, finally enters into wave coupled (W) mode in various configurations [20, 21]. Sometimes, the discharge directly transits from E mode into W mode. With RF power or magnetic field increase further, the wave coupled (W) mode may exhibit a centralized blue core (BC) or big blue, characterized by a concentrated column of blue light along the discharge tube [22, 23]. The Ar I (atom) and Ar II (ion) spectral lines are mainly in the 740–900 nm and 360–530 nm ranges. The visible light emissions mainly shifted from Ar I into Ar Ⅱ when discharge enters into BC mode. The plasma density and ionization rate are significantly higher in the blue core region than those outside of the core region. This phenomenon has been studied by researchers in various helicon plasma apparatuses under specific conditions [24-27]. The blue core is also considered a new transition mode after the traditional wave mode [28].

A typical helicon plasma system comprises a coupled antenna to excite the helicon wave and electromagnet apparatus to produce the magnetic field. The conventional antennas in experiments are Nagoya Ⅲ, Boswell, Half Helix, and Loop. The Nagoya III antennas have been demonstrated to be particularly effective in plasma-wave interactions, and as a result, they are widely used in industry. The Nagoya Ⅲ antenna consists of two rings connected by straight wires on both sides. The Boswell antenna, also referred to as a double-saddle antenna, features a circular ring on either side, divided into two semicircular segments. It represents a modified version of the Nagoya III antenna, and it was the antenna used on the experimental device that first experimentally reported helicon waves supported discharges. The Half Helix antenna represents a further improvement on the Nagoya III antenna. In this configuration, the straight wire of the Nagoya III antenna is twisted into a half helix. A loop antenna is one of the simplest loop structures. The loop antenna is azimuthally symmetrical. Chenwen Wang *et al* studied the effect of inhomogeneous magnetic field on blue core in Ar helicon plasma, using the Half Helix antenna in experiments [29]. Michael D. West *et al* employed the Boswell antenna to generate the plasma and observed a high-density "blue" mode in Xenon [30, 31]. Guilu Zhang *et al* obtained the maximum Ar+ ion flux of $7.8 \times 10^{23}$ m$^{-2}$ s$^{-1}$ with a bright blue core employing the Nagoya Ⅲ antenna [32]. Antar *et al* used a double-loop antenna in experiments and found that as the plasma remains in the "blue mode", increasing magnetic field leads to an increase in the density but a decrease in the electron temperature. The increase in neutral pressure yields the same effects [33]. J. E. Stevensa *et al* characterized a new helicon plasma source with a four-turn flat spiral coil in 1995 [34]. S. Shinohara *et al* studied the characteristics in large-diameter plasma produced by a planar spiral antenna, and their results show the successful excitation and the propagating helicon wave with azimuthal mode number $m = 0$ [35]. They also demonstrated that employing spiral antenna could increase the plasma diameter effectively [36]. Many other studies were also conducted using this planar spiral antenna to explore physics of plasma discharge, power absorption, wave structure, and applications of material treatment and space propulsion [37-52]. Recently, a novel planar antenna with four concentric rings was developed and tested on a linear plasma experiment, i.e. Linear Experimental Advanced Device (LEAD) [53]. This antenna can work in form of both a spiral antenna by connecting inner and outer rings and a concentric antenna by applying various directional currents to independent rings. The latter is particularly useful to introduce radially layered current drive which is highly desirable for studying azimuthal instabilities, core-edge mode coupling, and radial transport barrier [12].

This work aims to explore the detailed physics of blue-core helicon discharge using this antenna, i.e. in terms of spatial and temporal evolutions. We will show that the produced blue-core plasma driven by this antenna is indeed different from that driven by other antennas, i.e. hollow profile of plasma density in radius. We also find that the plasma is rotating in discharge driven by this antenna. And a rotating plasma can possibly drive a centrifugal instability. Then we employ two-fluid flowing plasma model to study centrifugal instability. This research is expected to contribute to a deeper comprehension of the formation process of blue-core helicon plasma, thus holding significance for the scholarly community.

The subsequent sections are organized as follows. In section 2, a detailed overview of the experimental apparatus configuration and diagnostic equipments are provided. In section 3, the experimental findings regarding mode transittions, centrifugal instability, and spatial and temporal evolutions, along with subsequent discussions, are presented. Finally, in section 4, we draw some conclusions on the spatial and temporal evolutions of blue-core helicon plasma, notably that plasma density peaks off axis.

## 2. Experiments

The investigation of the LEAD device is focused on exploring key physical problems related to plasma transport and turbulence, helicon discharge physics, and interactions between plasma and materials. Additionally, the device serves as a testing ground for a new diagnostic system designed for use in Tokamak devices.

The experiment is performed in a linear plasma device named LEAD, which is shown in Fig. 1 [53]. It consists of a cylindrical nonmagnetic stainless steel discharge tube of 0.4 m inner diameter and 1.5 m length, including the source chamber for plasma excitation and diagnostic chamber for plasma physics research. The target chamber is for plasma-material interaction (PMI), and its inner diameter and length are 0.9 m and 1.2 m, respectively. These chambers are surrounded by 15 magnetic coils of varying diameters. It is convenient to adjust the current in coils to generate the required magnetic field configuration and strength, ranging from 0 to 2000 G. The working gas in this experiment is argon, and the flow rate is controlled by the CS1000 mass flow controller, which can vary from 0 to 1 SLM in increments of 0.01 SLM. The vacuum pump systme mainly consists of two molecular pumps and mechanical pumps. And the molecular pump is capable of reaching speeds of up to 24,000 revolutions per minute. The working pressure can be maintained at 0.5–3 Pa by a vacuum pump system for plasma discharge.

Figure 2 illustrates the configuration of the employed antenna, comprising four interconnected circular copper RF antennas arranged concentrically. The diameters of the rings are 0.32, 0.24, 0.16, and 0.08 m, respectively. Each ring is equipped with a cut at the 12 o'clock position to facilitate the feeding of the antenna. Each ring can be connected by a copper belt through the cut, which can be easily changed to modify antenna configuration, particularly in order to alter the current direction in each ring. Moreover, the antenna is surrounded by an RF shield. In our experiments, four rings are employed and connected in series with each other, therefore forming a spiral antenna. And the current in each ring is in the same direction. The antenna is located at about z=0 m. A cooling water system is employed to facilitate thermal reduction, with the power supply, matching box, antenna, and chiller all connected to this system. The frequency of RF power supply is 13.56 MHz, and RF matching network is auto-matching $\pi$ type keeping reflected power less than 1% of input power.

The Langmuir probe, which is capable of moving in the radial direction, is positioned at window 3 and located about z=1.2 m to diagnose the primary parameters of plasma. The probe features a cylindrical tungsten probe tip that is 1.2 mm in diameter and 2 mm in length. The cylindrical probe is oriented perpendicular to the applied magnetic field. The minimum distance of probe movement can be lowered to 1 mm to measure the detailed spatial distribution of plasma. The probe voltage $V$ and current $I$ signals are recorded by a data acquisition system. The data acquisition system with a sampling rate of 1 M/s is employed to produce the $V$-$I$ curve. The biasing voltage sweeps from -30 V to 30 V at frequency of 100 Hz. And the plasma density and electron temperature can be caculated from the $V$-$I$ curve. To obtain spatial profiles of the plsama parameters, steady-state discharge is mantained while the probe sacans radially with an electric linear stage. For each radial position, the signal sampling duration is 1 s to allow the biasing voltage sweeps 100 times. An averaged $V$-$I$ curve is produced to derive the mean plasma parameters. In RF plasmas, the issue of RF interference arises. This phenomenon distorts the measured $I$-$V$ curve due to the plasma potential and the sheath potential oscillating at RF frequencies. This may result in the calculated electron temperature being higher than it would otherwise be. The probe is situated up to 1.2 m sufficiently away from the antenna, the impact of the RF source on the probe is already weak. Consequently, no discernible distortions are observed on the $I$-$V$ curve.

An ICCD (intensified charge-coupled device) camera is used to record the discharge images. The progression over time of the helicon wave discharge was documented using a Phantom v2012 high-speed camera equipped with a CMOS (Complementary Metal-Oxide-Semiconductor) sensor measuring 35.8 × 22.4 mm. The camera has a resolution of 1280 × 800 pixels, a maximum frame rate of 22,500 frames per second (fps). The high-speed camera is located at about z=3.0 m. More detailed information about LEAD can be found in previous studies [53, 54].

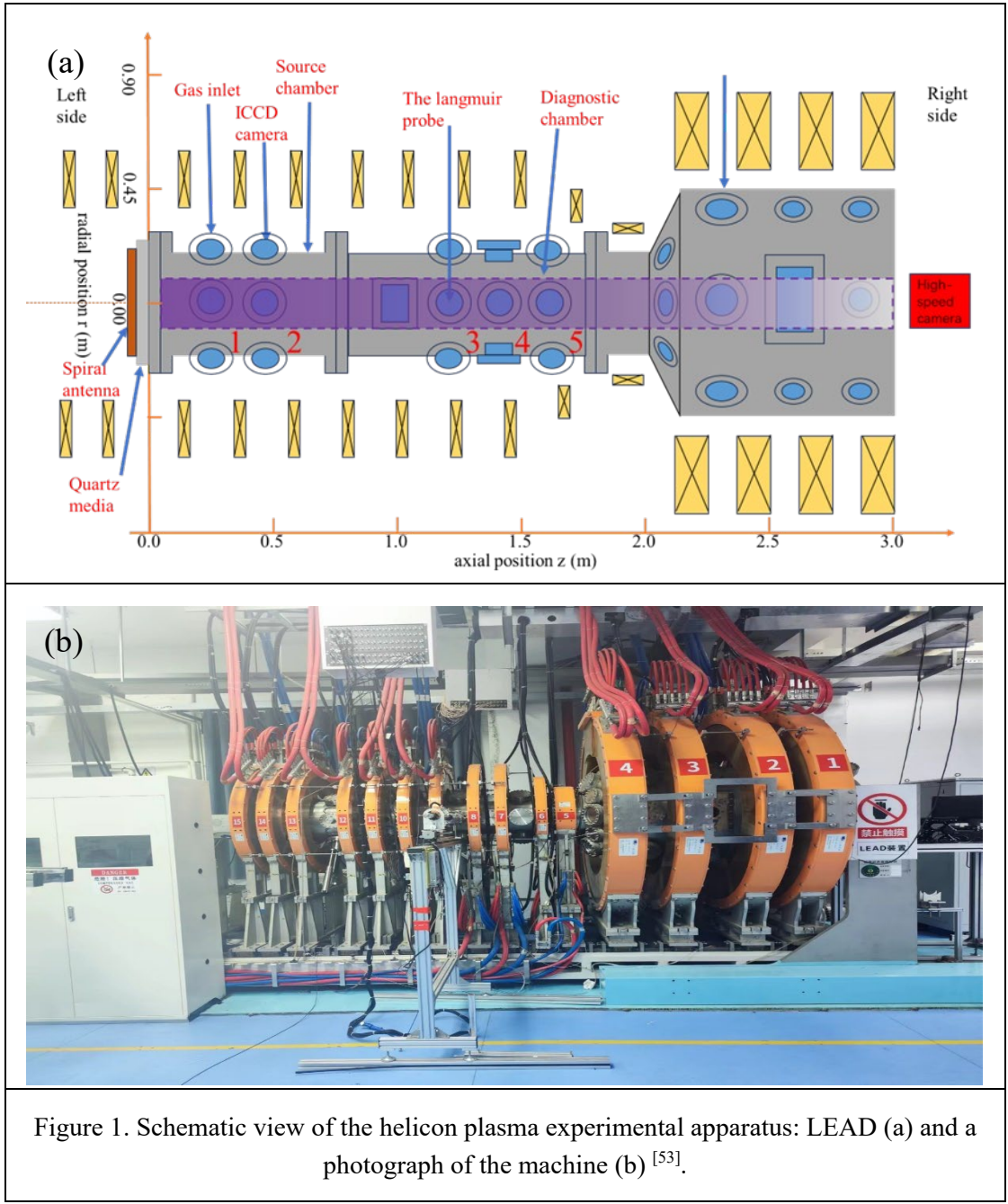

Figure 1. Schematic view of the helicon plasma experimental apparatus: LEAD (a) and a photograph of the machine (b) [53].

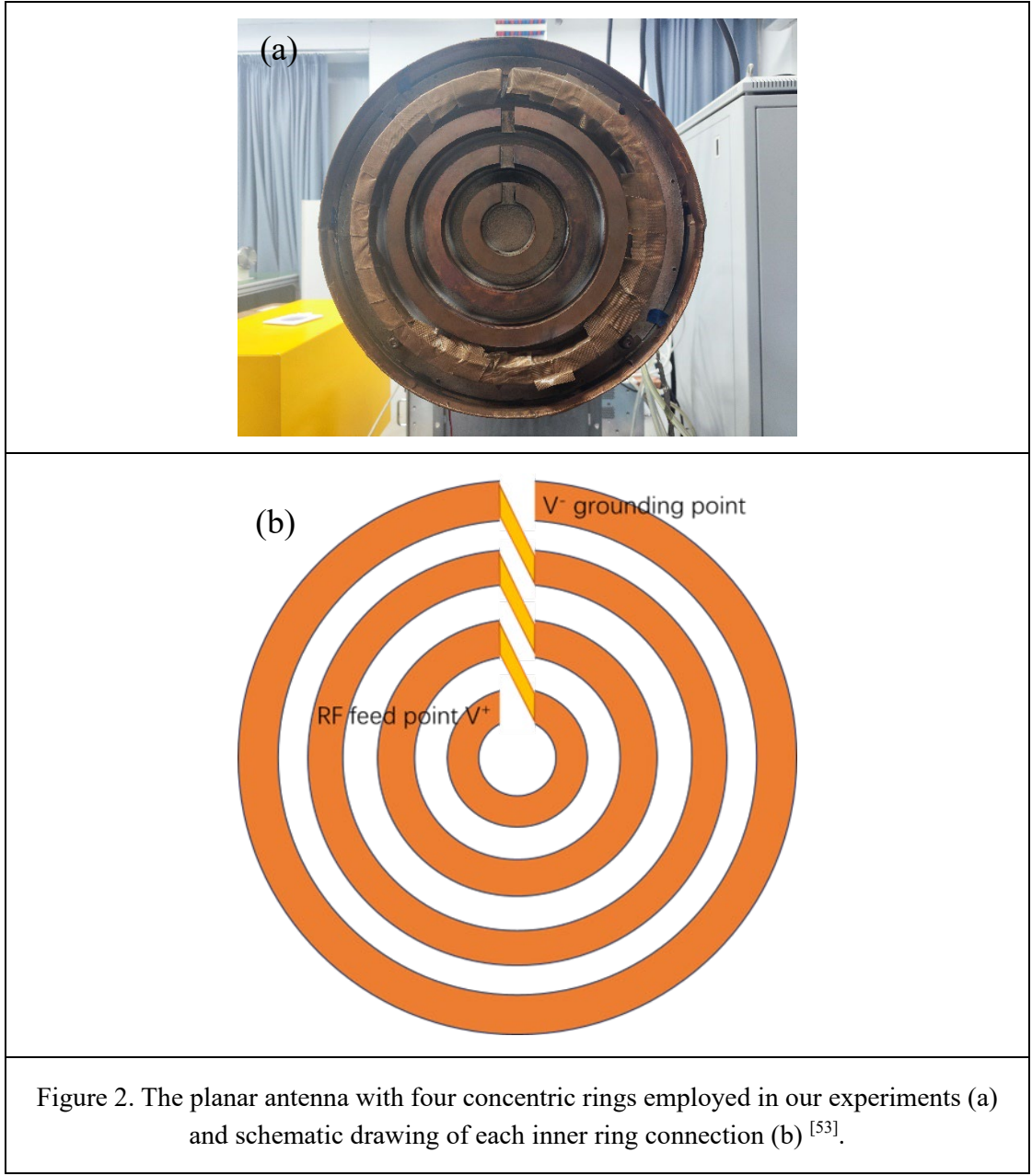

Figure 2. The planar antenna with four concentric rings employed in our experiments (a)
and schematic drawing of each inner ring connection (b) [53].

## 3. Experimental results and discussion

### 3.1 *Mode transitions and blue-core appearance*

First, power scanning experiments were carried out to explore the variations of plasma density at different input power levels in the device, and the results are displayed in Fig. 3. In this work, the electron density is used to express plasma density by assuming the plasma to be quasi-neutral ($N_i \approx N_e$), where $N_e$ is electron density and $N_i$ is ion density. Different flow rates show different gas pressure levels. The employed 1.2 Pa of argon neutral pressure corresponds to a flow rate of 0.4 SLM. The magnetic field strength of 500 G is applied. During experiments these parameters are constant through magnetic coils and vacuum pump systmes.

Discontinuous density jumps can distinguish the discharge mode. As shown in Fig. 3, when the power is lower than 800 W, the plasma density has two density jumps, corresponding to three discharge modes, i.e. E mode, H mode, and W mode, respectively [18, 55]. Specifically, when the discharge goes into W mode for power above 800 W and further increases power to 1100 W, the discharge transits into blue-core mode, along with the light radiated from the plasma changing from pink to blue color, which is typically accompanied by an increase in neutral depletion and ionization [56]. We use BC1, BC2, and BC3 to represent different levels of blue-core mode. It also shows that the threshold power for blue-core mode is approximately 1100 W. The discharge occurs in BC1, and the blue light emissions initially combine with some pink light. Subsequently, the power increases further to 1500 W with discharge in the BC2. The pink light nearly disappears entirely, and blue light emissions fill the tube. When the input power exceeds 2200 W with discharge in the BC3, the plasma density remains relatively constant, indicating that the plasma density has reached a saturation value.

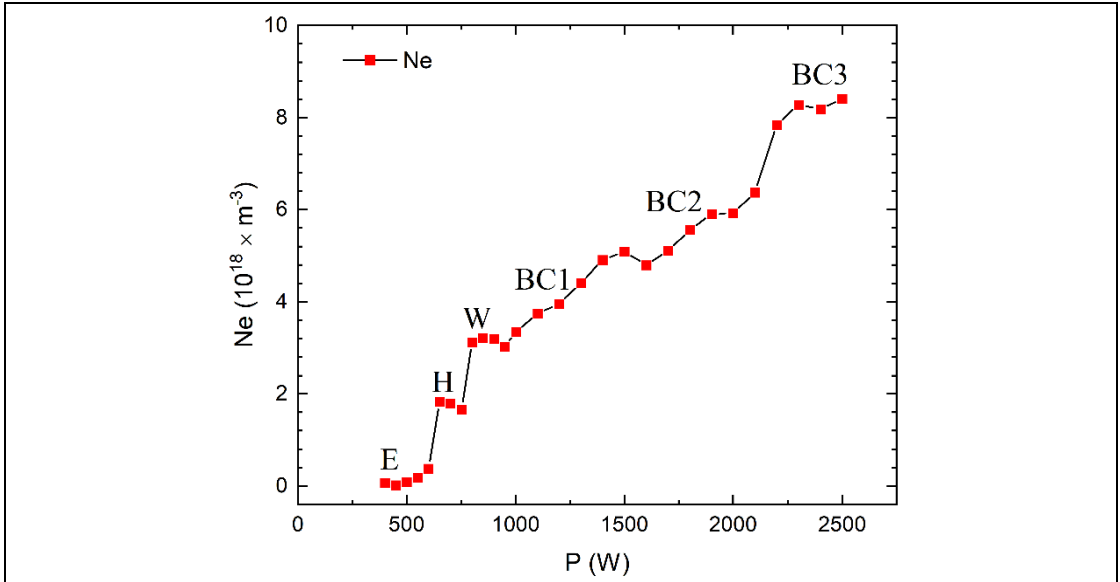

Figure 3. Discharge mode transitions: E, H, W, BC1, BC2, BC3 mode, respectively, with increased RF power for a flow rate of 0.4 SLM and magnetic field strength of 500 G. The Langmuir probe is placed at r = 0 cm to measure the displayed plasma density.

Subsequently, a high-speed camera was employed to document the temporal evolution of the discharge. As illustrated in Fig. 4, the discharge is in the non-blue-core mode for 500 and 1000 W, and it transitions into the blue-core mode for 2000 W. It is obvious that the pattern undergoes significant changes. To illustrate the evolution of the discharge in greater detail, as shown in Fig. 5, an increase in the input power from 500 W to 1000 W results in a shift from an annular to a fan-shaped light pattern, accompanied by an increase in light intensity. The rotational frequency of the 1000 W is about 90 rad/s and is also greater than that of the 500 W. The rotation could be explained by an $E_r$ x $B_z$ drift, for which $E_r$ is the electric field formed by the density gradient of plasma and $B_z$ is the external magnetic field. As the power increases, the electric field gradient becomes more pronounced, resulting in a stronger drift and faster rotation. Moreover, in the case of non-blue-core mode, the $E_r$ is positive, resulting in a clockwise rotation.

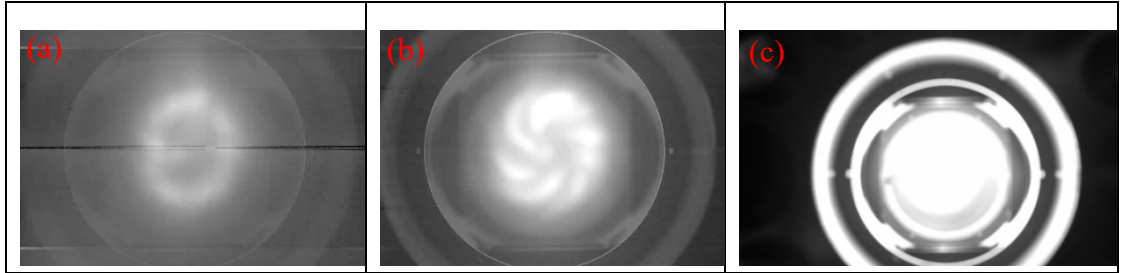

Figure 4. Images of discharge at the conditions of 500 G, and a flow rate of 0.4 SLM. The input power is (a) 500 W, (b) 1000 W, and (c) 2000 W, respectively. The speed, exposition time, and resolution are 1000 fps, 900 μs, and 960×540 px, respectively.

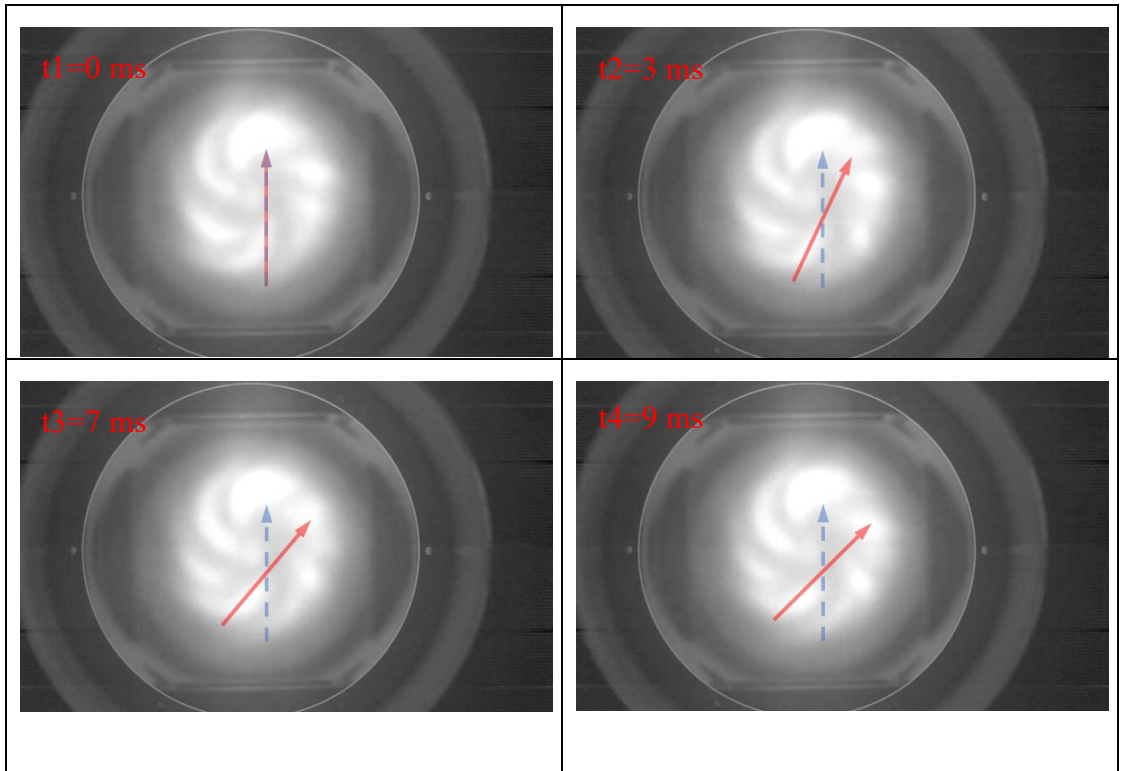

Figure 5. Images of discharge at the conditions of 500 G, 1000 W, and a flow rate of 0.4 SLM. The blue arrow is fixed, meaning the direction of reference, and the red arrow means the direction of rotation. The speed, exposition time, and resolution are 1000 fps, 900 μs, and 960×540 px, respectively.

However, as the discharge transitions into blue-core mode for 2000 W, as shown in Fig. 6, the blue light appears and the blue column forms. Moreover, the bright column rotates in a counterclockwise direction, diverging from the non-blue-core mode. The rotational frequency of 2000 W is 800 rad/s, and it is also augmented in comparison to that of 1000 W. The plasma potential and radial electric field demonstrated in the Fig.7, the plasma potential at 2000 W exhibits an inflection point at r = 5 cm, while the plasma potential at 1000 W demonstrates two inflection points, situated at r = 2 cm and r = 5 cm, respectively. The radial electric field is obtained through the application of a linear fit of the plasma potential and the derivative with respect to the radial distance. It is found that the radial electric field amplitude is lower than that in the non-blue-core mode when the discharge enters the blue-core mode. The radial electric field direction undergoes two changes in the non-blue-core mode and once in the blue-core mode, which may lead to a difference in the rotational direction between the two modes.

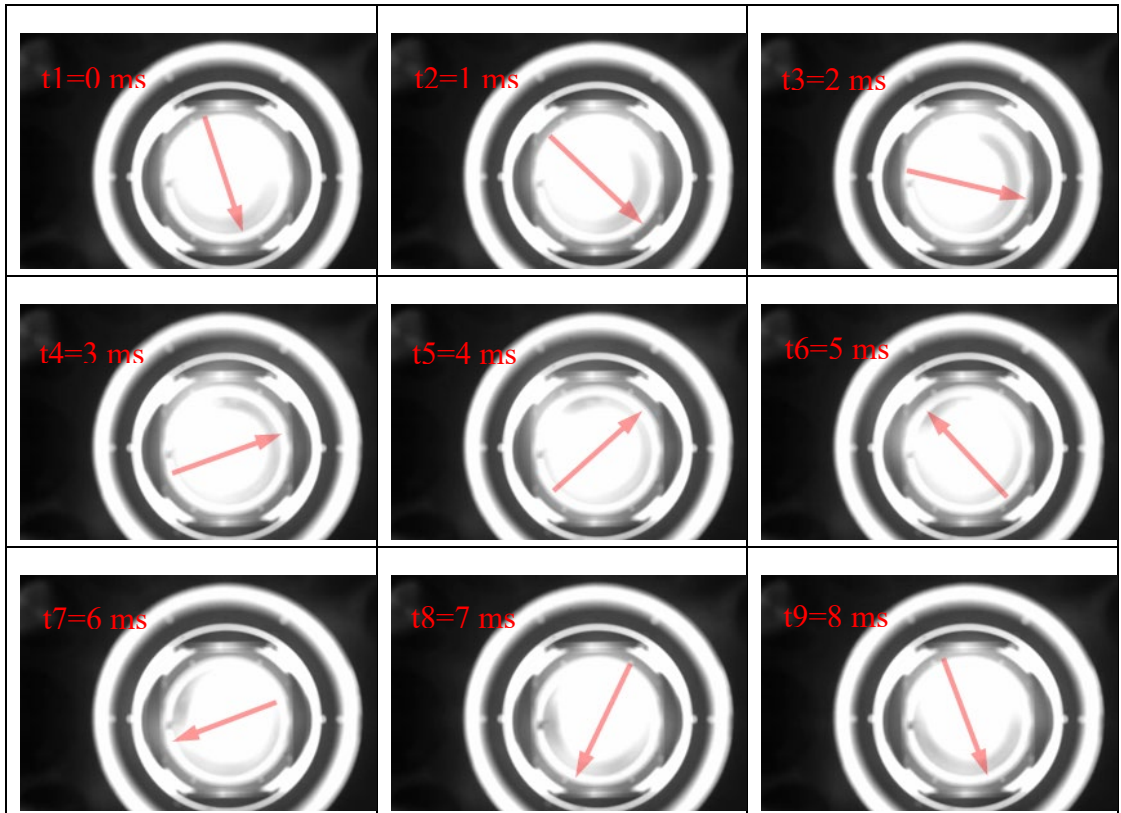

Figure 6. Images of discharge at the conditions of 500 G, 2000 W, and a flow rate of 0.4 SLM. The speed, exposition time, and resolution are 1000 fps, 900 μs, and 960×540 px, respectively.

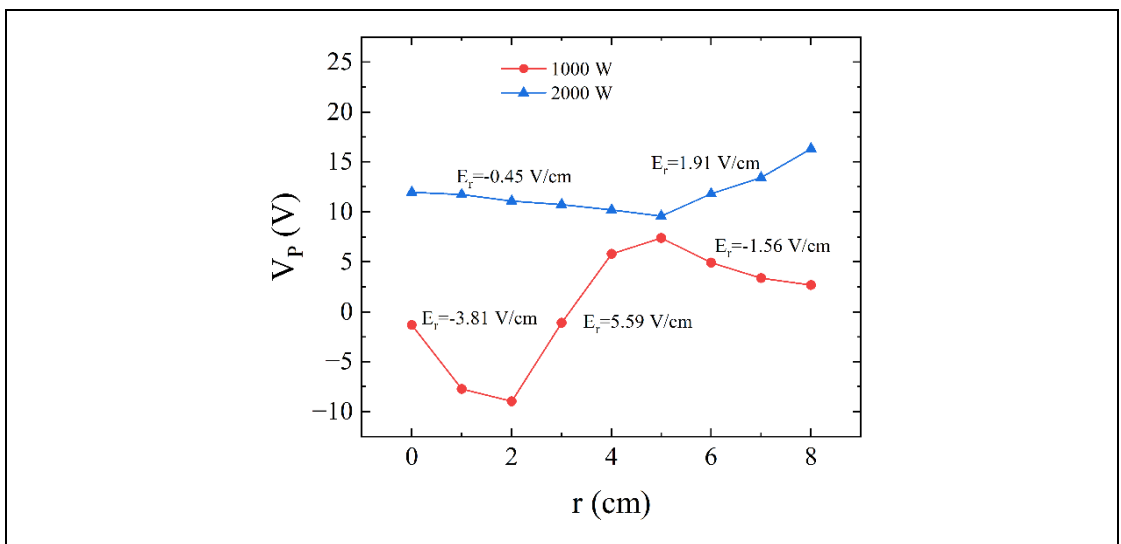

Figure 7. Radial profiles of plasma potential and corresponding radial electric field,

respectively.

To optimize the discharge conditions for blue-core mode, aiming for higher plasma density, and stable discharge, various flow rates and magnetic field strengths are employed. As shown in Fig. 8(a), it is obvious that there is a significant plasma density difference for flow rates of 0.6 to 0.8 SLM, respectively, spanning the magnetic field strength range from 500 to 1000 G. It is clear that plasma density decreases with increasing flow rates when magnetic field strengths are larger than 600 G. It is possible that the neutral pressure is too high to ionize adequately, which would result in a reduction in plasma density. However, for relatively low magnetic field strengths, plasma density shows a max value at a flow rate of 0.6 SLM and a magnetic field strength of 600 G. Moreover, plasma density also nearly decreases with increasing magnetic field strength for flow rates above 0.6 SLM. A similar phenomenon was also seen in a previous study [57].

A profile of electron temperature with various magnetic field strengths and flow rates is shown in Fig. 8(b). Similar to plasma density, electron temperature also shows obvious differences for flow rates of 0.7 and 0.8 SLM: As the magnetic field strength increases, the electron temperature gap between 0.7 and 0.8 SLM flow rates becomes larger, especially at high magnetic fields of 700 and 800G. Moreover, the electron temperature gradient for various magnetic field strengths becomes larger with decreasing flow rates. It is found that that electron temperature is increasing with an increasing applied magnetic field strength. This phenomenon can be attributed to the observed reduction in plasma density, which coincides with a decrease in the collision frequencies between electrons and neutrons. Consequently, the loss of energy experienced by electrons is diminished, resulting in an elevated electron temperature.

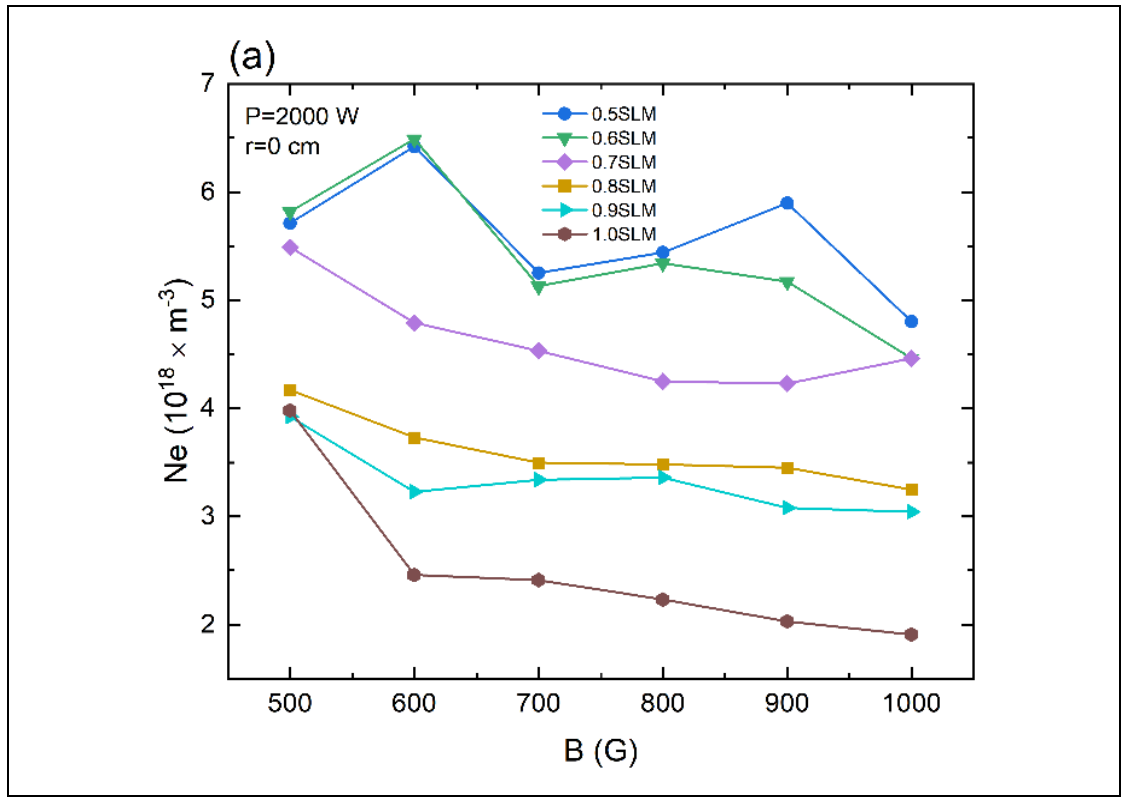

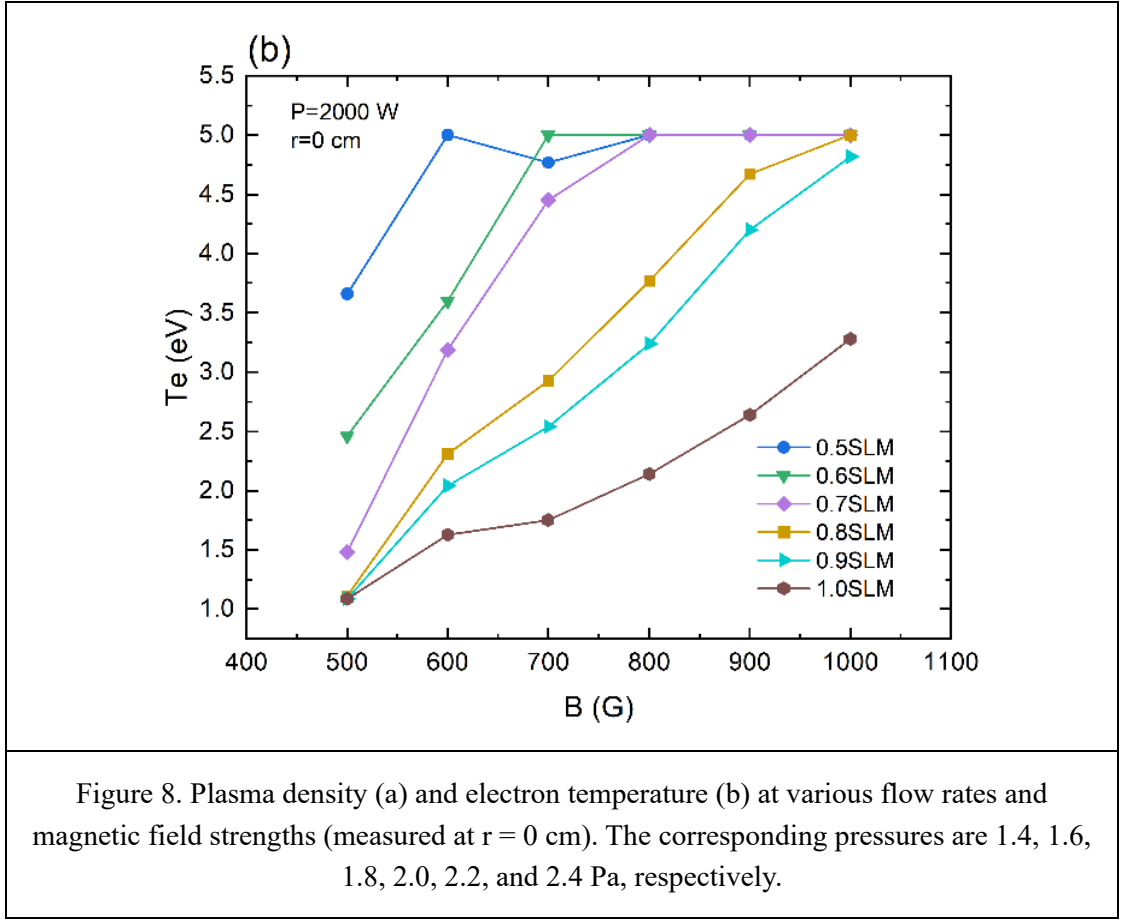

Figure 8. Plasma density (a) and electron temperature (b) at various flow rates and magnetic field strengths (measured at r = 0 cm). The corresponding pressures are 1.4, 1.6, 1.8, 2.0, 2.2, and 2.4 Pa, respectively.

As shown in Fig. 9, a camera was positioned at window 2 to capture images of the plasma discharge at various flow rates. The camera settings are ISO (international organization for standardization, sensitivity of the negative to light): 50, exposure compensation: 0, aperture value: f/1.9, focus: 24 mm, and white balance: auto white balance. The exposure times are 1/268, 1/193, 1/157, 1/136, 1/119, and 1/102 s, respectively, corresponding to Fig. 7 (a) to (f). Images were captured with a 2000 W input power and 500 G magnetic field strength. A blue column located at the regime center is seen, spanning the gas flow rates from 0.5 to 1.0 SLM. At 0.5 and 0.6 SLM, however, the blue emissions are intense and fill the entire tube. The blue emissions get weaker, and the pink light starts to show and gets stronger as the gas flow rate rises above 0.7 SLM. The blue light emission corresponds to the intense ion line emissions. The radial boundary area shows a pink annular color, corresponding to the intense atom line emission when the gas flow rate rises above 0.7 SLM [58]. Assuming a uniform radial electron temperature ($T_e$), the line intensity of atoms $I_{atom}$ and ions $I_{ion}$ has the following relationship [59]:

$$I_{atom} \propto n_0 n_e,$$
$$I_{ion} \propto n_e^2. \tag{1}$$

Here, $n_0$ and $n_e$ represent the neutral atom density and plasma density, while $I_{atom}$ and $I_{ion}$ refer to the line intensities of atoms and ions. As indicated in Eq. (6), the line intensity of the ion is the strongest for 0.5 and 0.6 SLM, and the plasma density is nearly the greatest. As the gas flow rate increases further, plasma density decreases, and neutral density increases, the line intensity of atoms becomes stronger. The variations in light emissions are in alignment with the plasma density variations from Fig. 8(a).

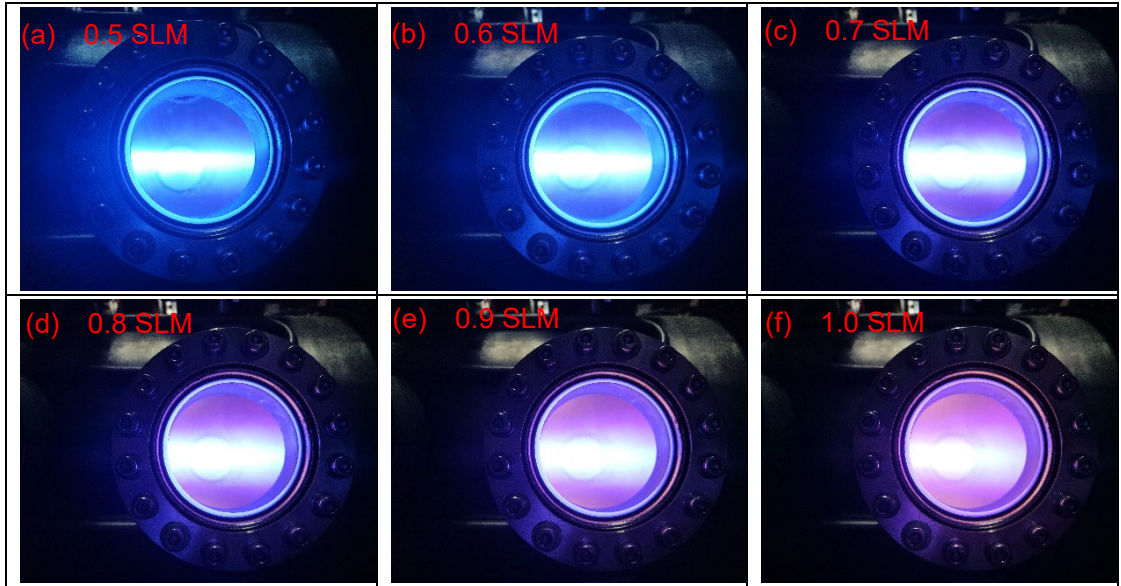

Figure 9. Light emission for different flow rates at the magnetic field strength of 500 G and input power of 2000 W observed from a side port on the small chamber stage: (a) 0.5 SLM, (b) 0.6 SLM, (c) 0.7 SLM, (d) 0.8 SLM, (e) 0.9 SLM, and (f) 1.0 SLM, respectively. The corresponding pressures are 1.4, 1.6, 1.8, 2.0, 2.2, and 2.4 Pa, respectively.

### 3.2 *The effect of RF power on radial configuration of blue core*

The camera settings are ISO (international organization for standardization, sensitivity of the negative to light): 50, exposure compensation: 0, aperture value: f/1.9, focus: 24 mm, and white balance: auto white balance. The exposure times are 1/100, 1/119, and 1/234, respectively, corresponding to Fig. 10 (a) to (c). As shown in Fig. 10, helicon discharge images at various input power are captured. It can be observed that with the increasing input power, the blue column gets brighter, and the purple light around the tube gets weaker and finally disappears. This phenomenon indicates that the ionization rate improves greatly. Next, we explore the effects of RF power on radial distribution. The discharge mode transits into blue-core mode at 1200, 1500, and 2000 W. The radial profile of plasma density is shown in Fig. 11. Previous studies [28, 60] showed that the plasma density is almost axially symmetric. Therefore, we consider only the positive radial direction of radial profiles. It can be seen that plasma density increases as input power increases. It's probable that increasing the power supply improves the power absorption of plasma, resulting in higher plasma density. Moreover, the plasma density peaks off axis when input power is 1500 and 2000 W showing a hollow shape. This phenomenon is significantly different from other typic antennas [28, 61]. It might be that the structure of the antenna is different from other linear devices. Specially, near the core region, the plasma density of 1200 W varies little with radius compared to that of 1500 and 2000 W. We also find that the trend of plasma density for 1200 and 2000 W is similar. However, for 1500 W, the gradient of plasma density is the largest. It indicates that power absorption may initially shrink onto the axis with input power, but as it further increases, the power absorption at the edge of the core region improves significantly, resulting in a smoother plasma density gradient.

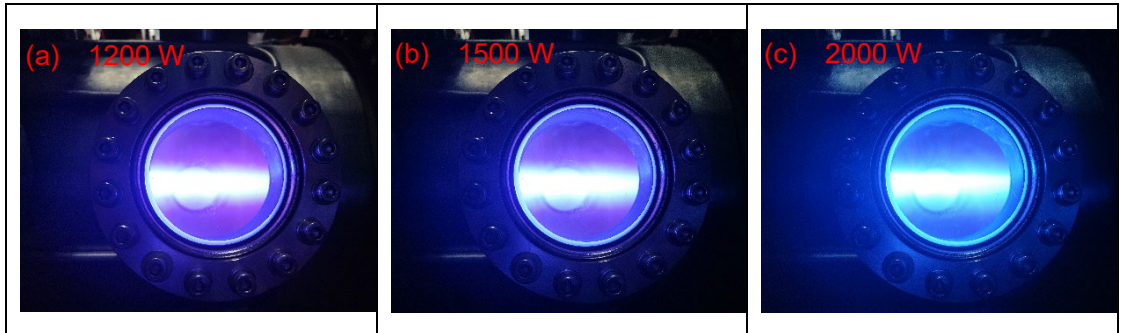

Figure 10. Light emission for different input powers at the magnetic field strength of 600 G and gas flow rate of 0.6 SLM observed from a side port on the small chamber stage: (a) 1200 W, (b) 1500 W, and (c) 2000 W, respectively.

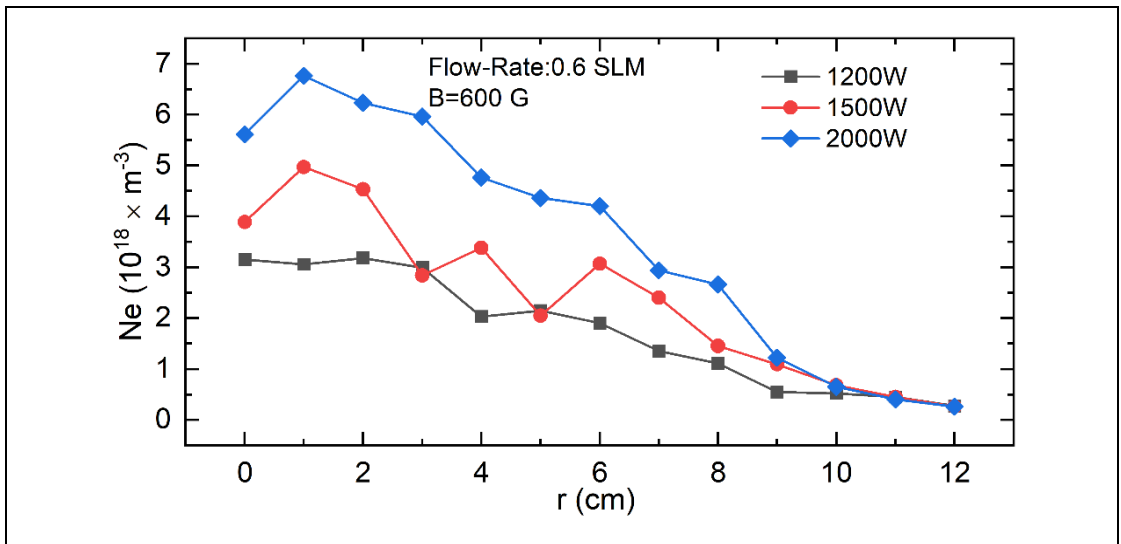

Figure 11. Radial profiles of plasma density with different input powers of 1200, 1500, and 2000 W, respectively. The magnetic field strength and flow rate are 600 G and 0.6 SLM.

Figure 12 shows the radial distribution of electron temperature under the same conditions as Fig. 11. It is interesting to note that the electron temperature displays divergent characteristics in comparison with the plasma density. In the core region of r < 4 cm, the electron temperature exhibits greater variability at 1200 W. It is noteworthy that the electron temperature distribution exhibits an overall "W" configuration, and the radial gradient of the electron temperature diminishes with an augmentation in the input power, a phenomenon that contrasts with the trend observed in the plasma density gradient. Moreover, in the core region of r < 4 cm, electron temperature shows the trend of shrinking onto axis with increasing input power. It may be indicating that the input power can enhance power absorption and intensify energetic electron collisions, resulting in a decrease in electron temperature as energetic electrons lose energy. This phenomenon is particularly evident in the core region.

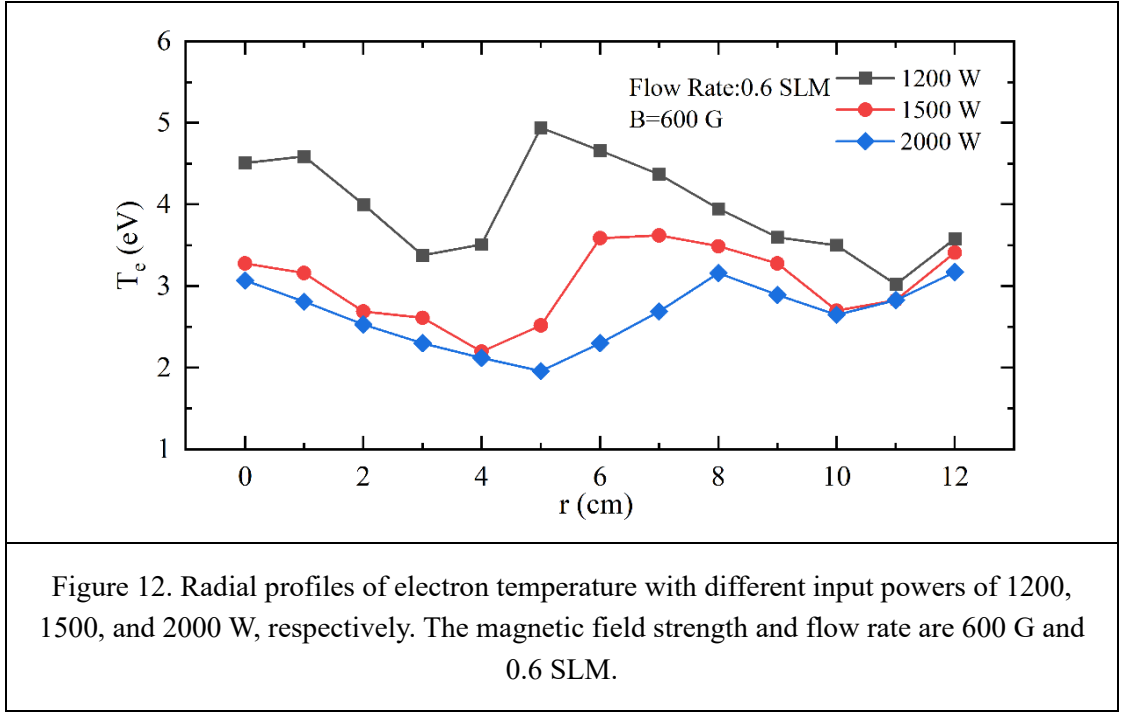

Figure 12. Radial profiles of electron temperature with different input powers of 1200, 1500, and 2000 W, respectively. The magnetic field strength and flow rate are 600 G and 0.6 SLM.

### 3.3 *The effect of flow rates on radial configuration of blue core*

Flow rate is a crucial parameter on plasma. The gas flow rates directly can influence the working pressure. Therefore, a flow rate scan is conducted to examine how the flow rate influences the radial distribution of the blue-core helicon plasma. As shown in Fig. 13, we first consider the radial distribution of plasm density with various gas flow rates of 0.5, 0.6, and 0.7 SLM, respectively. The magnetic field strength and input power are 600 G and 2000 W. It is obvious that the plasma density peaks off axis for 0.6 SLM. It also can be seen that plasma density is significantly larger in the radial distance at the gas flow rate of 0.6 SLM compared to that of flow rates of 0.5 and 0.7 SLM. Specifically, the maximum value of plasma density is about $6.8 \times 10^{18}$ m$^{-3}$ when the gas flow rate is 0.6 SLM, while the peak densities of 0.5 and 0.7 SLM are $4.6 \times 10^{18}$ and $4.8 \times 10^{18}$ m$^{-3}$. This indicates that a gas flow rate of 0.6 SLM is a better choice for blue-core helicon plasma with the present configuration and operating parameters. It also indicates that increasing the flow rate is not always helpful to increase the plasma density once the flow rate exceeds a threshold value. Increasing the gas inlet flow rate can enhance plasma density to a certain degree. This is likely due to a substantial increase in the number of neutral particles, which subsequently elevates the effective collision frequency between energetic electrons and these particles, thereby contributing to an increase in plasma density. It is also observed that plasma density for 0.5 and 0.7 SLM is nearly the same in the core region of 0 < r < 4 cm, while out of the core region (region of r > 4 cm), plasma density for 0.5 SLM becomes larger compared to that of 0.7 SLM. As gas flow rates increase, it is possible that neutral depletion become stronger and ionisation become weaker outside the core region (r > 4 cm), leading to a decrease in plasma density away from the core.

We can also see that the variations in plasma density are greater, and the radial gradient is roughly steeper at a gas flow rate of 0.7 SLM; however, it shows a relatively small variation in plasma density at the radial distribution at a gas flow rate of 0.6 SLM even though its discharge performs better in plasma density.

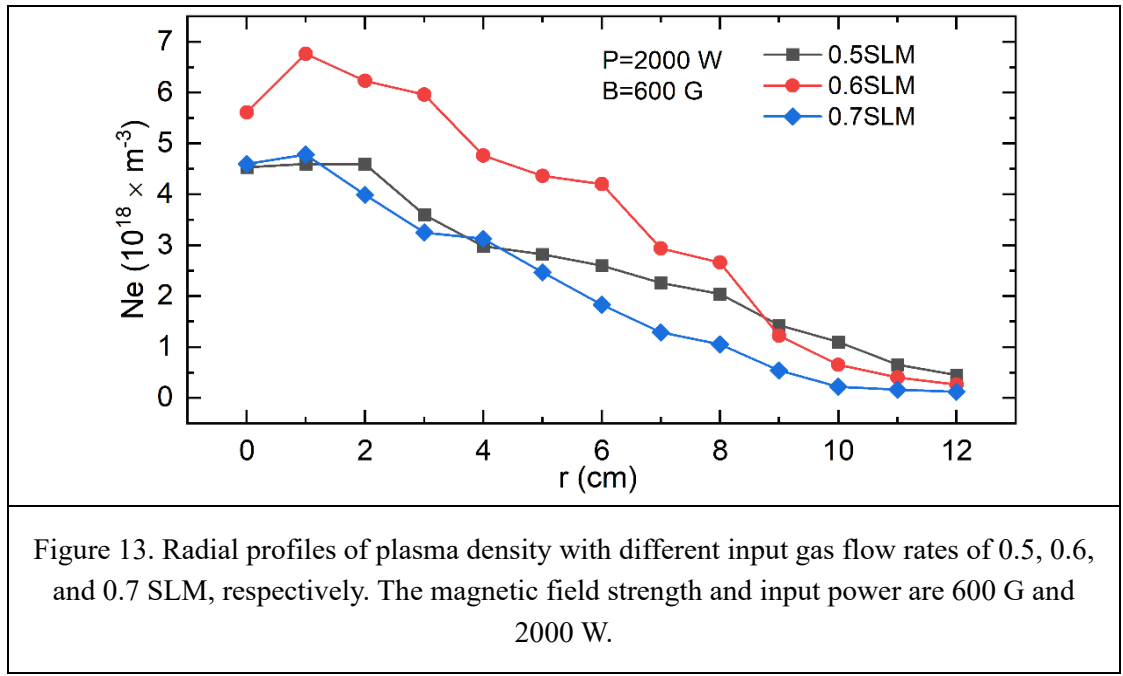

Figure 13. Radial profiles of plasma density with different input gas flow rates of 0.5, 0.6, and 0.7 SLM, respectively. The magnetic field strength and input power are 600 G and 2000 W.

Figure 14 shows the radial distribution of electron temperature under the same conditions as Fig. 13. It is found that electron temperature decreases with increasing gas flow rate. This phenomenon may be attributed to the fact that the number of neutral particles has increased significantly, resulting in an elevated collision frequency between electrons and neutral particles. This results in electrons losing a significant amount of energy, which subsequently leads to a reduction in electron temperature. Furthermore, near the core region, there is a larger difference in electron temperature between 0.5 and 0.6 SLM compared to that between 0.6 and 0.7 SLM. we can also see that the gradient of electron temperature at a gas flow rate of 0.6 SLM is larger near the core region. Moreover, the various trends of the electron temperature with gas flow rate show the opposite for the plasma density with gas flow rate. As illustrated in Fig.13, the plasma densities in the core region for 0.5 and 0.7 SLM are nearly identical. However, the electron temperature illustrated in Fig. 14 for 0.7 SLM is markedly lower than that observed for 0.5 SLM. This suggests that the decline in electron temperature for 0.7 SLM is not primarily caused by plasma density and is related to the increase in the number of neutral particles.

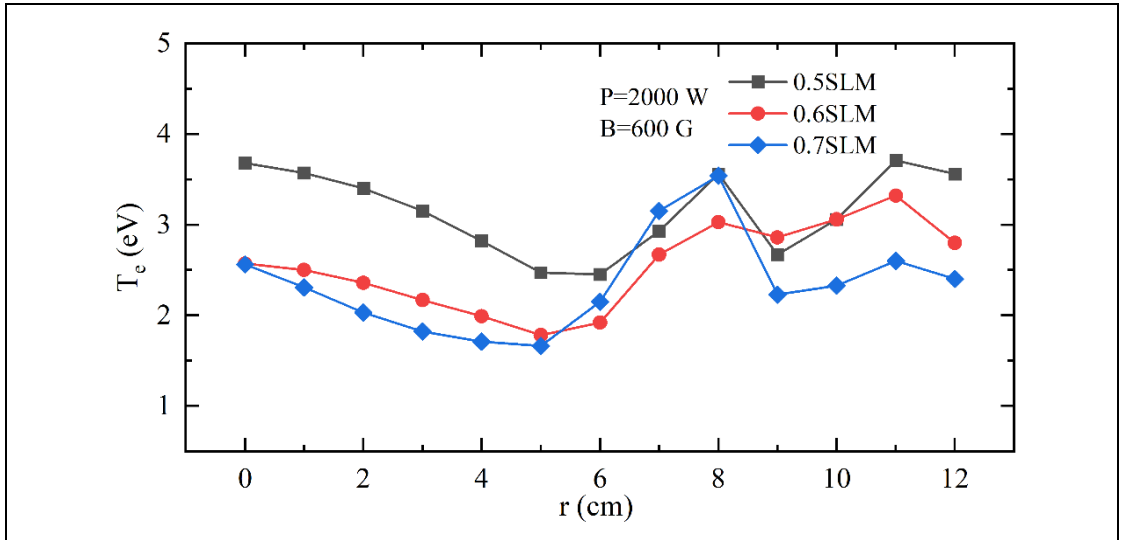

Figure 14. Radial profiles of electron temperature with different input gas flow rates of 0.5, 0.6, and 0.7 SLM, respectively. The magnetic field strength and input power are 600 G and 2000 W.

### 3.4 *The effect of magnetic field on radial configuration of blue core*

Next, we investigate the impact of magnetic field strength on the blue-core mode. Figure 15 displays the radial distribution of plasma density for different magnetic field strengths. The results indicate a significant decrease in plasma density between 800 and 1000 G and become flat in the core region, suggesting that the blue-core mode weakens when the magnetic field strength exceeds a threshold value of about 800 G. Additionally, at relatively low magnetic fields from 600 G to 800 G, the plasma density's radial distribution at 600 G is not steeper than that at 700 and 800 G, though the plasma density of 600 G is the largest among them. The results also show that the gradient of plasma density becomes steeper as the magnetic increases for 600, 700, and 800 G, aligning with the previous study [12]. Furthermore, the plasma density also peaks off axis among the magnetic field strengths of 600, 700, and 800 G, respectively.

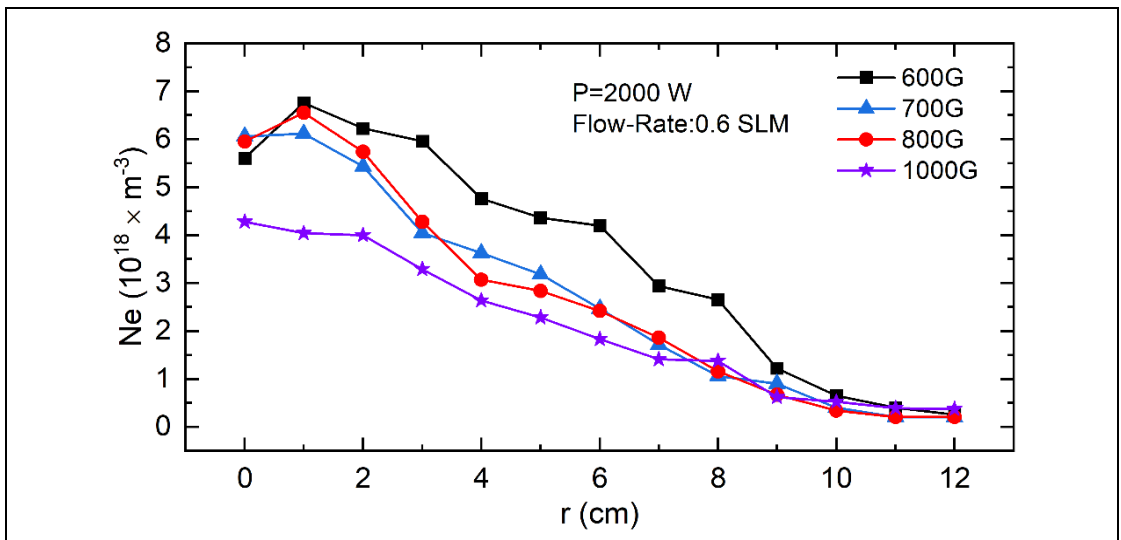

Figure 15. Radial profiles of plasma density with different magnetic field strengths of 600,

Figure 16 depicts the radial distribution of electron temperature under the same conditions as shown in Fig. 15. As demonstrated in the Fig. 16, in the region of the core, the electron temperature exhibits a reduced variation along the radial direction and is, in general, lower than approximately 5 eV when the magnetic field strengths are 600 G, 700 G, and 800 G. At 1000 G, the electron temperature remains at around 7.5 eV. Furthermore, an increase in magnetic field strength is accompanied by an increase in electron temperature. This phenomenon can be attributed to the restriction of electron motion and reduction of electron-neutral particle collisions by an increased magnetic field strength. This, in turn, leads to a decrease in electron energy loss and an increase in electron temperature. This phenomenon is in contrast to the results observed in Fig. 4.12, yet is consistent with the phenomenon depicted in Fig. 4.8(b).

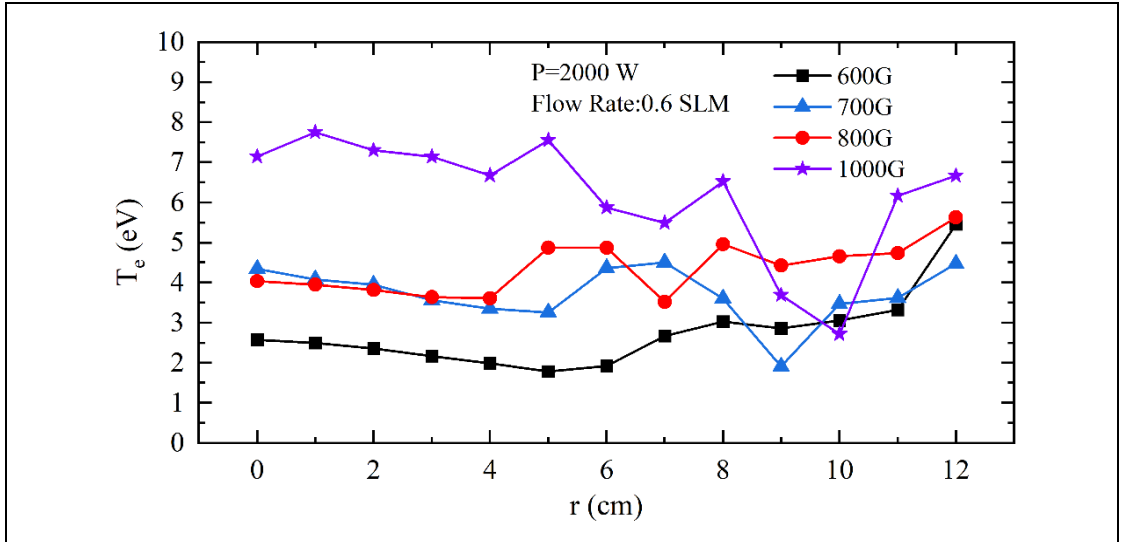

Figure 16. Radial profiles of electron temperature with different magnetic field strengths of 600, 700, 800, and 1000 G, respectively. The flow rate and input power are 0.6 SLM and 2000 W.

## 4.  Numerical Simulation

Since the rotation of plasma column will induce centrifugal instability, we shall calculate the rotation frequency from the video frames and then employ two-fluid flowing plasma model [62, 63] to compute the rotation induced centrifugal instability, which was not presented before for blue-core helicon plasma to our knowledge. The model employs following assumptions:

- Ions of different charge can be treated as a single species with average charge Z.

- The plasma is quasineutral, in the form of  $n_e = Zn_i$ .

- A steady-state plasma is characterized by azimuthal symmetry and the absence of an axial structure.

- Fluctuations in the externally applied magnetic field induced by plasma are not considered.

- The effects of the finite Larmor radius (FLR) and viscosity are not considered.

- . Uniform ion and electron temperature, $T_i$ and $T_e$ across the plasma column.

- The neglect of electron inertia is reasonable for the range of frequencies considered in this work.

- The steady-state ion density distribution is given by the Gaussian profile: $n_0 = n_i(0)e^{-(r/R)^2}$, where $n_i(0)$ is the on-axis ion density and $R$ is the characteristic radius at which the density is $1/e$ of its on-axis value.

- The steady-state velocities of ions and electrons can be expressed as $\mathbf{v_i} = (0, \omega_i r, v_{iz})$ and $\mathbf{v_e} = (0, \omega_e(r)r, v_{ez}(r))$, respectively. In these expressions, $\omega_i$ is the ion rigid rotor rotation frequency, $v_{iz}$ is the ion uniform axial streaming velocity, $\omega_e(r)$ is the electron rotation frequency, and $v_{ez}(r)$ is the electron streaming velocity. Radial diffusion of both ions and electrons due to electron–ion collision is not considered.

In this work, length and time are normalised to $R$ and $1/\omega_{ic}$, respectively. Here, $\omega_{ic} = ZeB_z/m_i$ is the ion cyclotron frequency. A cylindrical coordinate system is developed, with the following definitions: $(x, \theta, \zeta) = (r/R, \theta, z/R)$ and $\tau = \omega_{ic}t$, where $x$ and $\zeta$ are the normalised radial and axial positions, respectively.

The model includes the momentum and continuity equations for ion and electron fluids, respectively:

$$\frac{\partial \mathbf{u_i}}{\partial \tau} + (\mathbf{u_i} \cdot \nabla)\mathbf{u_i} = -\psi(Z\nabla\chi + \lambda\nabla l_i) + \mathbf{u_i} \times \hat{\zeta} + \delta n_s \tilde{\tilde{\xi}} \cdot (\mathbf{u_e} - \mathbf{u_i}), \tag{2}$$

$$\psi Z(-\nabla l_i + \nabla\chi) - \mathbf{u_e} \times \hat{\zeta} + \delta n_s \tilde{\tilde{\xi}} \cdot (\mathbf{u_i} - \mathbf{u_e}) = 0, \tag{3}$$

$$-\frac{\partial l_i}{\partial \tau} = \nabla \cdot \mathbf{u_i} + \mathbf{u_i} \cdot \nabla l_i, \tag{4}$$

$$-\frac{\partial l_i}{\partial \tau} = \nabla \cdot \mathbf{u_e} + \mathbf{u_e} \cdot \nabla l_i. \tag{5}$$

The terms are defined as follows:

$$\mathbf{u_i} = \frac{\mathbf{v_i}}{\omega_{ic}R} = (x\varphi_i, x\Omega_i, u_{i\varsigma}), \mathbf{u_e} = \frac{\mathbf{v_e}}{\omega_{ic}R} = (x\varphi_e, x\Omega_e, u_{i\varsigma}),$$

$$\lambda = \frac{T_i}{T_e}, \psi = \frac{k_B T_e}{m_i \omega_{ic}^2 R^2},$$

$$\chi = \frac{e\phi}{k_B T}, \tilde{\xi} = diag(\xi_\perp, \xi_\perp, 1),$$

$$l_i = \ln \frac{n_i}{n_i(0)}, n_s = \frac{\nu_{ei}}{\nu_{ei}(0)}, \delta = \frac{eZn_{i0}}{B_z}\frac{\eta_L}{\gamma_E}.$$

Here, the subscripts $i$ and $e$ denote ion and electron parameters, respectively. The $\varphi$ represents the normalized radial velocity divided by x, while $\Omega$ denotes the normalized rotation frequency. The $u_\varsigma$ indicates the normalized axial velocity. The $\lambda$ is defined as the ratio between ion and electron temperatures. The constant $\psi$ becomes the square of the normalized ion thermal velocity for $\lambda = 1$. The $\chi$ is a normalized electric potential $\phi$, $l_i$ is the logarithm of the ratio of the ion density $n_i$ to its on-axis value $n_i(0)$, and $n_s$ is the ratio of the electron–ion collision frequency $\nu_{ei}$ to its on-axis value $\nu_{ei}(0)$. Additionally, $\delta$ represents the normalized resistivity parallel to the magnetic field. Here, $\eta_L$ denotes the electrical resistivity of a Lorentz gas, while $\gamma_E$ signifies the ratio of the conductivity of a charge state Z to that observed in a Lorentz gas [64].

For large axial wavelength modes of the resistive plasma column, which is the case for our study, this model can be reduced to a second-order differential equation as follows:

$$(\frac{\varpi^2 - C^2}{\varpi \Psi})L(N_c)[g_1(y)] = 0, \qquad (6)$$

where

$$L(N_c) = y\frac{\partial^2}{\partial y^2} + (1-y)\frac{\partial}{\partial y} + (\frac{N_c}{2} - \frac{m^2}{4y}),$$

$$N_c = \frac{(\varpi^2 - C^2)(m + \frac{i}{2}f(y))}{\varpi - m\Omega_{i0}^2 + i\Psi f(y)} + \frac{mC}{\varpi},$$

and $f(y) = F^2 e^y$ with the normalized axial wave number $F = k_\varsigma / \sqrt{\delta}$. Here, we define

$\varpi = \omega - m\Omega_{i0} - k_\zeta u_{\zeta 0}$     that is the frequency in the frame of ion fluid,

$C = 1 + 2\Omega_{i0}$, $\Psi = (\lambda + Z)\psi$. Equation 5 are $g_1(0) = 0$ and $g_1(Y) = 0$ with the infinite

radius $Y$ representing the edge of plasma column. For even $m$, these conditions become

$g_1^{'}(0) = 0$ and $g_1(Y) = 0$. The typical parameters employed in computation are listed in Tab.

1 and get under the conditions of 2000 W, 500 G, and 0.4 SLM. Here, $\omega_0$ is rotation frequency of plasma, $V_{z0}$ is axial streaming velocity, $m_i$ is the argon ion mass, and $B_z$ is the magnetic field strength.

Table 1 Parameters employed in computation

| Parameters | Value |
|:---:|:---:|
| $n_i$ (on axis) | $6 \times 10^{18}$ m$^{-3}$ |
| $T_e$ | 5 eV |
| $T_i$ | 0.8 eV [65] |
| $m_i$ | 40 amu (Ar) |
| $B_z$ | 0.05 T |
| $Z$ | 1.0 |
| $V_{z0}$ | 200 m/s |
| $\omega_0$ | 0.8 krad/s |
| $\omega_{ic}$ | 120 krad/s |
| $\Omega_{i0} = \dfrac{\omega_0}{\omega_{ic}}$ | 0.0066 |
| $\Psi = (\dfrac{T_i}{T_e} + Z)\dfrac{k_B T_e}{m_i \omega_{ic}^2 R^2}$ | 9.65 |
| $\delta = \dfrac{eZn_{i0}}{B_z} \dfrac{\eta_L}{\gamma_E}$ | 0.0037 |
| $R$ | 0.08 m |

The dispersion curves for the plasma conditions of LEAD are shown in Fig.17. For blue-

core helicon plasmas of LEAD, the peak of normalized growth rate $\varpi^i = 0.0122$ lies at

$F \approx 0.11$, with normalized frequency $\varpi^r = 0.064$: the wave increases the fastest in the frame

of the ion fluid. However, $F > 0.11$, the normalized growth rate $\varpi^i$ decreases gradually and

reaches approximately 0.005 when $F \approx 1$. Moreover, it can be seen that normalized frequency

$\varpi^r$ increases sharply when $F < 0.11$ while $F > 0.11$, $\varpi^r$ increases slowly as $F$. This may indicate that centrifugal instability decreases as increasing $F$.

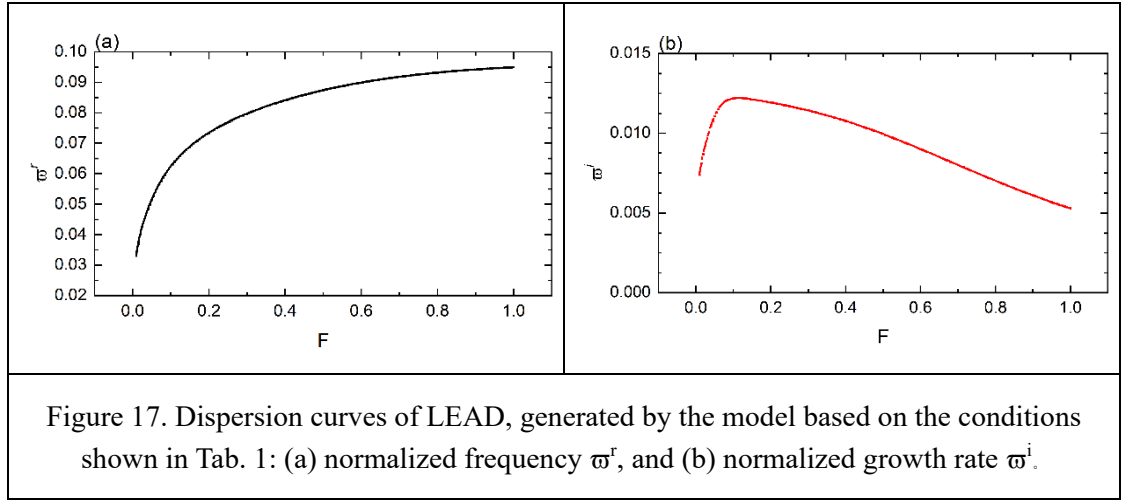

Figure 17. Dispersion curves of LEAD, generated by the model based on the conditions shown in Tab. 1: (a) normalized frequency $\varpi^r$, and (b) normalized growth rate $\varpi^i$.

Figure 18 illustrates the radial profiles of the perturbed density, $n_{i1}$, in conjunction with the equilibrium density gradient, $|n_{i0}'(r)|$. The data was computed here under the conditions $B_z = 0.05$ T, $R = 0.08$ m, and $T_e = 5$ eV. It is observed that the perturbed density has one peak lying at approximately r=0.045 m. The equilibrium density gradient also has one peak at approximately 0.055 cm. The coincidence of their radial locations implies that the instability has resistive drift-type characteristics driven by density gradient [57]. Furthermore, the perturbed density and the equilibrium density gradient decrease sharply to 0 when $r$ is greater than 0.15 m. This may suggest that resistive drift occurs in the core region.

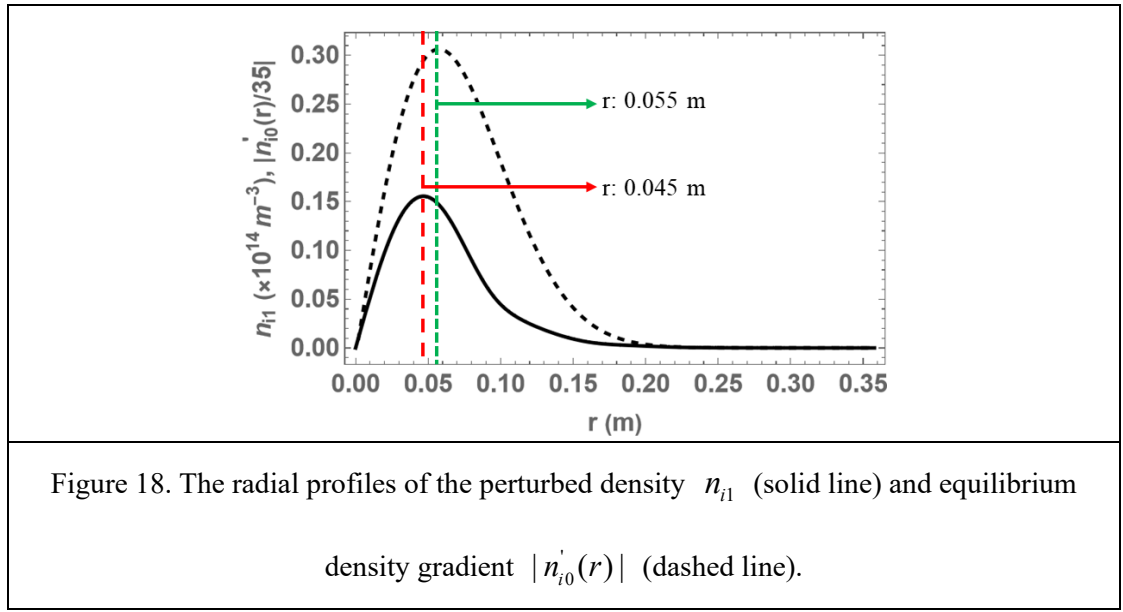

Figure 18. The radial profiles of the perturbed density $n_{i1}$ (solid line) and equilibrium density gradient $|n_{i0}'(r)|$ (dashed line).

## 5. Conclusion

Blue-core mode of helicon discharge has been attracting wide research interests because of various special phenomena involved. While most previous studies explore the phenomena using traditional helicon experiment setups, our work intends to fill the gap by studying a planar antenna with concentric rings. This setup is particularly useful for producing large-area helicon plasma. This work presents experimental measurements on the LEAD apparatus which employs planar antenna with four concentric rings that is connected by each ring to form a spiral antenna, and two-fluid flowing plasma computations, to study the spatial and temporal evolutions of blue-core helicon discharge in detail.

Important findings are summarized as follows.

- Under the conditions of a gas flow rate of 0.4 SLM and magnetic field strength of 500 G, the discharge undergoes mode transitions and finally converts into blue-core mode when the input power is above 1100 W;

- For the blue-core mode, the bright column rotates in the counterclockwise direction, and the rotation frequency is greater compared to that of the non-blue-core mode, and the underlying drive is attributed to $E_r \times B_z$ drive;

- The profile of plasma density peaks off axis for the blue-core mode, i.e. about r=1 cm, showing hollow structure in radial, which is confirmed by experimental measurements and direct observations through eyes.

- The electron temperature decreases as the input power increases, while the radial gradient of the electron temperature inside the core is smaller as the magnetic field changes;

- There exists a threshold value of inlet gas flow rate (i.e. 0.6 SLM in our case) for which plasma density maximize. This also applies to the electron temperature because of the varied collision frequency between electrons and neutral particles when changing the gas flow rate;

- The plasma density shrinks onto the axis as magnetic field strengths increase. However, once the magnetic field increases above 800 G in this experiment, the plasma density decreases significantly, and the blue-core becomes weak.

- From the two-fluid flowing plasma computations, it is found that for blue-core mode, the centrifugal instability becomes weak as the axial wave number increases, and the coincidence of radial locations of maximum perturbed density and equilibrium density gradient implies that the instability is a resistive drift mode driven by density gradient;

At present, more advanced diagnostic equipments, including the voltage-current probe, B-dot magnetic probe array [66], and local optical emission spectroscopy [67], are being developed and installed on the LEAD to characterize the blue-core helicon plasma with much more details. Future research will be devoted to further exploring the phenomena and physics of blue-core helicon discharge by varying the current direction on each ring of the antenna. The idea is trying to introduce radially layered current drive and match the radially enclosed blue-core column. The efficiency of power coupled from antenna into plasma will be analysed in situ via installing an impedance monitoring system. The detailed physics of transport barrier and sheared flow will be also computed by using a hybrid PIC-MHD code, combined with experimental data.

**Supplementary Material**

The supplementary material includes three high-speed camera recordings of plasma discharge under an external magnetic field of 500 G, an inlet flow rate of 0.4 SLM, and discharge power of 500 W, 1000 W, and 2000 W, respectively. These videos (Supplementary Videos S1–S3) complement Figs. 4–6, providing visual evidence of the plasma's temporal evolution and rotational dynamics.

## Acknowledgements

This work is supported by the National Natural Science Foundation of China (92271113, 12411540222, 12481540165), the Fundamental Research Funds for Central Universities (2022CDJQY-003), and the Chongqing Entrepreneurship and Innovation Support Program for Overseas Returnees (CX2022004).